\documentclass[12pt]{article}
\usepackage{amssymb}
\usepackage{amsmath}
\usepackage[ps,dvips,matrix,arrow,frame,import,curve,color]{xy}
\makeatletter
\@addtoreset{equation}{section}
\makeatother

\def\be{\begin{equation}}
\def\ee{\end{equation}}

\def\ba{\begin{array}}
\def\ea{\end{array}}

\newcommand{\bea}{\begin{eqnarray}}
\newcommand{\eea}{\end{eqnarray}}
\def\N{$\cal N$}

\def\E {$E_{7(7)}$}

\begin{document}
\hfill{}

\begin{flushright}
\end{flushright}

\vskip 1cm

\vspace{24pt}

\begin{center}
{ \LARGE {\bf    Light-by-Light Scattering Effect\\
\vskip 0.4cm

 in Light-Cone Supergraphs }}

\vspace{24pt}

{\large  {\bf     Renata Kallosh${}^\dag$ and Pierre Ramond${}^\ddag$ }}

    \vspace{15pt}

{${}^\dag$Stanford Institute of Theoretical Physics, Department of Physics,\\
 Stanford University, Stanford, CA 94305}

{${}^\ddag$ Institute for Fundamental Theory, Department of Physics, \\
 University of Florida, 
Gainesville FL 32611
 }


\vspace{10pt}

\vspace{24pt}

\end{center}

\begin{abstract}

We give a relatively simple explanation of the light-cone supergraph prediction  for the  UV properties of the maximally supersymmetric theories. It is based on the existence of a dynamical supersymmetry which is not manifest in the light-cone supergraphs. It suggests that 
\N=4 supersymmetric Yang-Mills theory is UV finite and \N=8 supergravity is UV finite at least  until 7 loops whereas the   $n$-point amplitudes have no UV divergences  at least until $L=n+3$. Here we show  that this prediction can be deduced from the properties of light-cone supergraphs analogous to the light-by-light scattering effect in QED. A technical aspect of the argument relies on the observation that the dynamical supersymmetry action is, in fact, a compensating field-dependent gauge transformation required for the retaining the light-cone gauge condition $A_+=0$.

\end{abstract}
\newpage
\newpage
\section{Introduction}

The light-cone superfield method for  \N=4 supersymmetric Yang-Mills theory and for \N=8 supergravity was proposed long time ago \cite{Brink:1982pd}, 
\cite{Mandelstam:1982cb}. It was used immediately to prove the UV finiteness of the \N=4 supersymmetric Yang-Mills theory in \cite{Mandelstam:1982cb}. The interest to this formalism returned during the next few years when it was discovered that \N=8 supergravity has better UV properties than expected \cite{Dixon:2010gz}, \cite{Bern:2009kd}.

Based on the properties of the linearized dynamical supersymmetry for the asymptotic superfields in a light-cone formalism a prediction was made in  \cite{Kallosh:2009db}  for the UV properties of the amplitudes. For  \N=4 supersymmetric Yang-Mills theory the argument gives an alternative proof of  UV finiteness,  for  and \N=8 supergravity it suggest that the theory is  UV finite at least  until 7 loops whereas the  $n$-point amplitudes have no UV divergences  at least until $L=n+3$. In particular the 4-point amplitude in \N=8 supergravity is UV finite at least  until 7 loops. This is a possible explanation of the 3-loop and 4-loop finiteness of the theory discovered in \cite{Bern:2009kd}. 

The argument in \cite{Kallosh:2009db} is based on the helicity formalism for the amplitudes and on the light-cone superfield formalism for  \N=4 supersymmetric Yang-Mills theory  and \N=8 supergravity.  Without a detailed knowledge of these two formalisms the UV prediction of \cite{Kallosh:2009db} is best understood by examining the relation between dynamical supersymmetry and gauge symmetries in the light-cone which will be given in this paper.

The light-cone superfield formalism starts with the component formalism where the gauge-symmetry is fixed in the light-cone gauge $A_+=0$ for YM theory (and $h_{+\nu}=0$ for supergravity). In this gauge the unphysical  $A_-$ and  $\psi_-$ fields are integrated out 
and the theory in components depends only on physical fields, helicity $\pm 1$ vectors $A_1 \pm i A_2$, spinors and scalars. 

The ensuing light-cone action can be rewritten using the scalar superfields which makes half of the original 16 supersymmetries manifest. These 8 supersymmetries are called kinematical, they form an algebra $\{\bar Q, Q \}=p_+$. The remaining supersymmetries are called dynamical and they are not manifest, in the same way as the Lorentz symmetry, which was broken from the start by the gauge condition $A_+=0$. 

The light-cone gauge condition $A_+=0$ is not sufficient to pin down the gauge symmetry\footnote{with thanks to D. Belyaev and W. Siegel}.  The existence of such a residual  gauge symmetry is best seen in terms of the dynamical supersymmetry transformations of the ${\cal N}=4$ theory. At the free level, they are linear in both the transverse derivatives and the superfields  of the theory.  Their generalization to superconformal interactions was shown in \cite{Ananth:2005zd} to be very simple: dynamical supersymmetry transformations become quadratic in the superfields, in such a way that the transverse derivatives are generalized to {\em covariant derivatives}. This indicates a residual gauge symmetry, which is the remnant of the original gauge symmetry with a gauge parameter independent of $x_-$. It is enough to suggest that amplitudes must depend on the transverse field strengths, and therefore have a different degree of divergence than naively expected. This situation is analogous to Delbr\"uck scattering, the scattering of light by light in QED, where a four-photon amplitude is actually a dimension eight (irrelevant) operator due to the derivatives acting on the external photon lines. 

As we proceed to show in detail, this argument serves to improve considerably the ultraviolet properties of both ${\cal N}=4$ and ${\cal N}=8$ maximally supersymmetric theories.  {\it On each external leg with the chiral superfield $\phi(p, \eta)$ there is an extra transverse momentum $p_\bot$ } due to linearized dynamical supersymmetries.  On vector fields they are, in fact, the compensating field-dependent gauge transformations required for the retaining the light-cone gauge condition $A_+=0$.

\section{Light-cone gauge $A_+=0$}

In  the Yang-Mills theory without supersymmetry the light-cone gauge $A_+=0$ does not fix the 
symmetry completely. We require that $A_+=0$ as well as $\delta_{\Lambda(x)}  A_+=0$
\be
\delta_{\Lambda(x)}  A_+  = \nabla_+  \Lambda(x)\equiv  (\partial_+ +  A_+) \Lambda(x) = {\partial \over \partial x_-} \Lambda(x_+, x_-, x_i) =0 \qquad \Rightarrow \qquad \Lambda(x_+, x_i) \ .
\label{x-}\ee
This condition can be solved by an $x_-$-independent gauge parameter $\Lambda(x_+, x_i)$, so that on the remaining gauge fields the partially local symmetry is still acting,  e. g. the remaining symmetry transformation on the transverse gauge fields $A_i$ is
\be
\delta_\Lambda A_i=\nabla_i \Lambda(x_+, x_i) =( \partial _i + A_i) \Lambda(x_+, x_i) \neq 0 \ .
\label{restr}\ee
Under the restricted gauge transformations (\ref{x-})
the theory is still invariant
since the gauge symmetry (\ref{restr}) with a gauge parameter depending on $x_+, x_i$ remains a partially local symmetry.

In a supersymmetric Yang-Mills theory the situation is somewhat different. Before gauge-fixing there is a local gauge transformation and a global susy transformation
\be
\delta_{ \Lambda (x) +\rm susy } \,  A_{\mu}(x) = \nabla_\mu \Lambda (x) + \bar \epsilon \gamma_\mu \psi(x) \ .
\label{cov1} \ee
One can gauge-fix the gauge $A_+=0$ and preserve this condition  so that $\delta_{\rm \Lambda + \rm susy}   A_+=0$ by performing a gauge transformation together with the susy transformation and requiring that
\be
\delta_{\rm \Lambda + \rm susy} \,  A_{+}(x) = \partial _+ \Lambda (x) + \bar \epsilon \gamma_+ \psi(x)=0
  \qquad \Rightarrow \qquad \Lambda( \bar \epsilon, \psi(x))= -  \bar \epsilon  \gamma_+  {1\over \partial_+} \psi(x) \ .
\label{susy}\ee
Thus the original gauge symmetry parameter $\Lambda(x)$ is not arbitrary anymore, it depends on a global susy parameter $\bar \epsilon$ and on the spinorial field $\psi(x)$, it is  denoted $\Lambda( \bar \epsilon, \psi(x))$ and it is given in eq. (\ref{susy}). On the transverse gauge fields  the transformations include this field-dependent non-linear  $\bar \epsilon$-dependent  transformation preserving the gauge $A_+=0$, together with the  original supersymmetry
\be
\delta_{\rm \Lambda(\bar \epsilon, \psi(x)) + \rm susy} \,  A_{i}(x) =- \bar \epsilon  \gamma_+  \left ({\nabla _i  \over \partial_+}  \psi(x)
\right ) + \bar \epsilon \gamma_i \psi(x) \ .
\ee
In the light-cone gauge, after the non-physical $A_-$ fields and  $ \psi_-$ are integrated out\footnote {Here the spinorial field $\psi= \psi_++ \psi_-$ is split into a part  $\psi_+$ which is preserved in the light-cone supermultiplet and a part $ \psi_-$ which is absent. }, one finds that the second term in these transformations presents a kinematical supersymmetry whereas the first one corresponds to a dynamical one. We explain the details below using the  two-component notation, which is convenient for the helicity formalism.

\subsection{On kinematical and dynamical supersymmetries}

Here we start with the Lorentz covariant (not gauge fixed) action of the \N=4 YM theory in usual space in components, depending on the vectors, spinors and scalars, 
$
{\cal L} = -{1\over 4} F_{\mu\nu}^2 +...
$.
The {\it linearized symmetry of the asymptotic Lorentz covariant vector fields} includes the linear part of the non-abelian  gauge symmetry as well as 16 supersymmetries. On the vector potentials the symmetries act as  shown in eq. (\ref{cov1}),
or, in the momentum space in 2-component notation
\be\boxed{
\delta_{\rm cov} \, A_{\alpha \dot \alpha}(p) = p_{\alpha \dot \alpha} \Lambda (p) + \bar \epsilon_{\dot \alpha A}  \psi_{ \alpha}^{ A}+ \bar \psi_{\dot \alpha A}  \epsilon_{ \alpha}^A} 
\label{cov} \ee
In particular,
\be
\delta_{\rm cov} \, A_{2 \dot 2}(p) = p_{2 \dot 2} \Lambda (p) + \bar \epsilon_{\dot 2 A }  \psi_ 2^{ A}+ \bar \psi_{\dot 2 A}  \epsilon_{2}^A
\ee
and \be
\delta_{\rm cov} \, A_{1 \dot 2}(p) = p_{1\dot 2} \Lambda (p) + \bar \epsilon_{\dot 2 A}  \psi_{ 1}^{ A}+ \bar \psi_{\dot 2 A}  \epsilon_{1}^A \ .
\ee
To fix the gauge $A_+=0$ we require that the supersymmetry transformation of the field $A_+$ is compensated by the field-dependent gauge transformation
\be
 \delta A_+=0  \quad \Rightarrow \quad \Lambda^{\rm comp} ( \psi, \bar \psi)= - {1\over p_+} \left ( \bar \epsilon_{\dot 2 A } \psi_{ 2}^{ A}+ \bar \psi_{\dot 2 A}  \epsilon_{ 2}^A\right)\equiv  {1\over p_+} \left(\Sigma^{\rm comp} (\psi) +\overline \Sigma ^{\rm comp} (\bar \psi)\right).
\label{comp}\ee
Thus the theory requires a compensating gauge transformation with the parameter which is a linear combination of the supersymmetry parameters and spinor fields.

This means that in the light-cone gauge $A_{2\dot 2}= A_+=0$ the physical vector fields  $A_{1\dot 2}$ and $A_{\dot 1 2}$  transform under supersymmetry as shown in eq. (\ref{cov}) where the compensating gauge transformation parameter  (\ref{comp}) has to be used. For example, 
\be
\delta_{\rm g.f.} A_{1 \dot 2}(p) = p_{1 \dot 2} \Lambda^{\rm comp} ( \psi, \bar \psi) + \bar \epsilon_{\dot 2 A}  \psi_{ 1}^{ A}+ \bar \psi_{\dot 2 A}  \epsilon_{ 1}^A \ .
\ee
This can be rewritten in the form 
\be
\delta_{\rm g.f.}  A_{1 \dot 2}(p) =   - {p_{1 \dot 2}\over p_+} \left ( \bar \epsilon_{\dot 2 A}  \psi_{ 2}^{ A}+ \bar \psi_{\dot 2 A}  \epsilon_{ 2}^A\right) + \bar \epsilon_{\dot 2 A}  \psi_{ 1}^{ A}+ \bar \psi_{\dot 2 A}  \epsilon_{ 1}^A \ .
\label{g.f.}\ee
Since we are looking at the symmetries of the free asymptotic fields, the fermions satisfy the Dirac equation
\be
p_{\dot 2 1} \psi_{2}^{ A} = p_{\dot 2 2 } \psi_{1}^{ A} \ .
\ee
This leads to a simplification of the gauge-fixed transformations of the vector field since the first and the third terms in eq. (\ref{g.f.}) cancel! The remaining two terms are
\be\boxed{
\delta_{\rm g.f.}  A_{1 \dot 2}(p) =   - {p_{1 \dot 2}\over p_+}  \, \bar \psi_{\dot 2 A}  \epsilon_{ 2}^A + \bar \psi_{\dot 2 A}  \epsilon_{ 1}^A=p_{1 \dot 2} \overline \Sigma ^{\rm comp} (\bar \psi) + \bar \psi_{\dot 2 A}  \epsilon_{ 1}^A} 
\label{g.f.1}\ee
The first term in this equation is given by a compensating gauge transformation and the second term we recognize as the kinematic supersymmetry.  For the conjugate field we find an analogous expression
\be\boxed{
\delta_{\rm g.f.}  A_{ \dot 1 2}(p) =    - {p_{ \dot 1 2}\over p_+}  \,  \bar \epsilon_{\dot 2 A } \psi_{ 2}^{ A}+ \epsilon_{ \dot 1 A}  \psi_{2}^{ A} =p_{ \dot 1 2}  \Sigma ^{\rm comp} ( \psi) + \epsilon_{ \dot 1 A}  \psi_{2}^{ A} }  
\label{g.f.con}\ee
Now we are ready to compare these transformations of the vector fields with those coming from the chiral scalar on shell superfield.

\section{Light-cone  path integral and symmetries}

 The  path integral for the generating functional of the on shell amplitudes studied in \cite{Kallosh:2009db},
  \cite{Fu:2010qi}
 is given by
 \be
 e^{i W[\phi_{in}(z) ]}= \int d\phi~  e^ {i {\cal S} [ \phi(z)] + i \int d^8 z\,   \phi_{in}(z) \, p^2\,  \phi(-z)} \ .
\label{pathintYM}\ee

Here for the \N=4 YM case  the  Lie-algebra valued off-shell superfield $\phi(p, \eta)= \phi^a(p, \eta)  t^a$ depends only on physical degrees of freedom of \N=4 SYM theory:
\bea
 \phi  = \bar A(p) + \eta_A \Psi ^A (p) + {1\over 2!} \eta_{A} \eta_B  \phi ^{AB}(p) + {1\over 3!} \epsilon ^{ABCD} \eta_{A}\eta_B \eta_C  \overline \Psi_D (p)+ {1\over 4!} \epsilon ^{ABCD} \eta_{A}\eta_B \eta_C  \eta_D  A(p) \ .
\label{PhiYM}\eea

The chiral scalar superfield $\phi(p, \eta)$, the integration variable in the path integral, is off shell, 
$
p^2 \phi(p, \eta)\neq 0$ whereas the asymptotic field  $ \phi_{in}(p, \eta)$ is on shell, $ p^2 \phi_{in}(p, \eta)=0
$.
In (\ref{pathintYM})
$z=(p, \eta)$ is  the 4+4 momentum superspace.  The integration is defined as
$
 d^8 z \equiv  {d^4 p\over (2\pi)^4} d^4 \eta
$.

For \N=8 supergravity the path integral is analogous, see the details in \cite{Kallosh:2009db}. The difference is that the chiral superfield is not Lie-algebra valued and the number of $\eta$'s is 8 instead of 4 as in \N=4 YM.

The linearized asymptotic symmetries of the free superfield $\phi_{in}$ are the following in \N=4 YM case.
There are 16 supersymmetries, $q^A_{\dot \alpha}= \bar \lambda_{\dot \alpha}{\partial \over \partial \eta_A}\, ,~  q_{B\alpha}=  \lambda_\alpha \eta_B$\, ,
\be
\boxed{
\delta \phi_{in}(p, \eta)= \left (\epsilon^{\alpha A} \lambda_\alpha \eta_A + \bar \epsilon^ {\dot \alpha}_A  \bar \lambda_{\dot \alpha}{\partial \over \partial \eta_A} \right)\phi_{in}(p, \eta  )}
\label{16}\ee
They form the closed the closed algebra
 \be
\{ \bar q^A_{\dot \alpha} , q_{B\alpha} \}=  \delta^A{}_B \, \lambda_\alpha \bar \lambda_{\dot \alpha } \ .
 \ee
In the light-cone formulation  8  
 supersymmetries  are manifest in the path integral, they are called  kinematical supersymmetries and are given by $q_{A 2}=\lambda_2 \eta_A$ and $  \bar q_{\dot 2}^A=\bar \lambda_{\dot 2}{\partial \over \partial \eta_A}$. They form the algebra
 \be
\{ \bar q^A_{\dot 2} , q_{B 2} \}=  \delta^A{}_B \, \lambda_2 \bar \lambda_{\dot 2 }=  \delta^A{}_B \, p_+ \, ,\qquad p_{2\dot 2 } = p_+ \, , \qquad  \lambda_2 = \bar \lambda_{\dot 2 }= \sqrt {p_+} \ .
 \label{kin}\ee 
The remaining 8 supersymmetries are the   
so-called dynamical supersymmetries given by  $q_{A 1}=\lambda_1 \eta_A$ and $  \bar q_{\dot 1}^A=\bar \lambda_{\dot 1}{\partial \over \partial \eta_A}$ and
 \be
\{ \bar q^A_{\dot 1} , q_{B 1} \}=  \delta^A{}_B \, \lambda_1 \bar \lambda_{\dot 1 }=  \delta^A{}_B \, p_- =\delta^A{}_B \, {p_\bot \bar p_{\bot}\over p_+}
\, , \qquad  \lambda_1 = {p_\bot \over  \sqrt {p_+}} \qquad ,  \lambda_{\dot 1} = {\bar p_\bot \over  \sqrt {p_+}} \ .
\label{dyn} \ee
The physical on shell amplitudes, computed via the supergraphs in eq. (\ref{pathintYM}) are expected to have all 16 linearized asymptotic symmetries. The prediction about the UV properties of the maximal supersymmetric QFT were made in \cite{Kallosh:2009db}  on the basis of  the non-manifest dynamical supersymmetry described above. 


\section{Covariant symmetries upon gauge-fixing versus light-cone superfield ones}
From the light-cone superfield transformations  ({\ref{16}) we find
\be
\delta \bar A=  \left (\bar \epsilon^ {\dot \alpha}_A  \bar \lambda_{\dot \alpha}{\partial \over \partial \eta_A} \right) \eta_B \Psi^B = \bar \epsilon^ {\dot \alpha}_A  \lambda_{\dot \alpha} \Psi^A \ ,
\ee
which means that
\be\boxed{
\delta \bar A=   - \bar \epsilon_ {\dot 2 A} \bar \lambda_{\dot 1} \Psi^A + \bar \epsilon_{\dot 1 A } \bar \lambda_{\dot 2} \Psi^A} 
\ee
Here the first term represent a dynamical supersymmetry with $\bar \epsilon^{\dot 1 }_A $ and the second one a kinematical one with $\bar \epsilon^{\dot 2 }_A $. For the conjugate vector field we have to compute
$
{1\over 4!} \epsilon ^{ABCD} \eta_{A}\eta_B \eta_C  \eta_D \delta  A=  \left ( \epsilon^ { \alpha A }  \lambda_{  \alpha}  \eta^A \right)  {1\over 3!} \epsilon ^{ABCD} \eta_{A}\eta_B \eta_C  \bar \Psi_D (p) $.
It follows that
\be\boxed{
 \delta  A= - \bar \Psi_A   \epsilon_ { 2}^{ A}  \lambda_{ 1} + \bar \epsilon_{\dot 1 A } \bar \lambda_{\dot 2} \Psi^A} 
\ee
We now see that with\footnote{This is in a precise agreement with spinor field rescaling suggested in  \cite{Kallosh:2009db} which is required to bring the original light-cone superfield of \cite{Brink:1982pd}, 
\cite{Mandelstam:1982cb} to the form given in eq. (\ref{PhiYM}).}
\be
\bar A= A_{\dot 1 2} \qquad    A= A_{1 \dot  2} \qquad \psi_{2}^{A} = \sqrt {p_+} \Psi
\ee
and the definition of the $\lambda, \bar \lambda$ in (\ref{kin}), (\ref{dyn}) we have identified all symmetries of the  space-time fields in a gauge-fixed theory with the symmetry transformations of the chiral light-cone superfield $\phi(p, \eta)$.

Via this identification we have also learned that the dynamical supersymmetry of  the light-cone superfield theory is actually a compensating gauge symmetry on the vector potentials preserving the gauge-fixing condition $A_+=0$:
\be\boxed{
 \delta_{\rm dyn} A(p) = \delta_{\rm gauge} A(p)  = p_\bot \overline \Sigma(\bar \psi)
\qquad \delta_{\rm dyn} \bar A(p) = \delta_{\rm gauge} \bar A(p)  = \bar p_\bot  \Sigma( \psi)} 
 \ee
 
\section {Implications for UV properties}

By observing that the dynamical supersymmetry in the on shell light-cone superfield is just a compensating gauge transformations we see the analogy with the well known concept in QED: scattering of light-by-light.
The UV divergences should not depend on  $A_\mu$, they should depend on the field strength $F_{\mu\nu}$. In gravity case they should not depend on $h_{\mu\nu}$ but on the linear part of $R_{\mu\nu\lambda\delta}$. This means that on each external leg $h_{\mu\lambda}$ there is an extra factor of $p_{\nu} p_{\delta}$ which leads to a better UV behavior of the Feynman graphs.

Note that in the covariant analysis of the UV divergences  the straightforward effect of ``scattering of light-by-light'' is already taken into account. The gravitational analog of this effect is the 3-loop 
candidate\footnote{This 3-loop $R^4$ counterterm for \N=8 supergravity, covariant and supersymmetric at the linear level was constructed long time ago in \cite{Kallosh:1980fi}. However, the computations in \cite{Dixon:2010gz,Bern:2009kd} have shown that the 3-loop  divergence is absent. Recently it was also noticed in 
 in the linearized covariant helicity formalism analysis of \cite{Elvang:2010jv} that  the candidate $R^4$ divergence is not ruled out. Therefore, the light-cone supergraph analysis of \cite{Kallosh:2009db}
gives  the only known explanation as to why in four dimensions the candidate $R^4$ divergence is absent, in agreement with \cite{Dixon:2010gz,Bern:2009kd}.}
divergence for the \N=8 supergravity, $R^4$. The fact that it depends on the space-time curvatures takes care of linearized gauge-invariance of the free graviton in the 4-point amplitude.

In the light-cone supergraph amplitudes the chiral superfield $\phi(p, \eta)$ is not a gauge-invariant object (not invariant under dynamical supersymmetry). The lack of gauge invariance is obvious, one can see from eq. (\ref{PhiYM}) that the superfield depends on gauge potentials $\bar A$ and $A$ and not on the field strength $p_\bot \bar A- \bar p_\bot A$.

The symmetry is restored when the external factors of transverse momenta are added for each external chiral superfield. For example, for the 4-point amplitude one finds \cite{Kallosh:2009db} that a special Grassmann $\delta$-function is required to secure the dynamical supersymmetry. In the \N=4 YM case
\be
A^{YM}(p_i, \eta_i) = \delta^4 \left (\sum_{i=1}^4\lambda_1^i \eta_i \right) {\cal P}^{YM}(p_i) \ .
\ee
Since $\lambda_1\sim p_\bot $ we see that for 4 chiral superfields  4 transverse momenta are necessary, see \cite{Kallosh:2009db} for details. These 4 transverse momenta mean that there is  a non-polynomial dependence on transverse momenta in the supergraph amplitude, by dimension. No non-polynomial dependence on the transverse momenta should appear in the UV divergences. This explains the UV finiteness of  \N=4 YM theory.
This is an  effect of  ``scattering of light-by-light'' in the supergraphs when some momenta are extracted from the internal lines as they  must be present on the external legs. 

In \N=8 supergravity the supergraph amplitude respecting  dynamical supersymmetry has an 8-dimensional $\delta$-function, which corresponds to adding 2 transverse momenta for each chiral superfield leg:
\be
A^{SG}(p_i, \eta_i) = \delta^8 \left (\sum_{i=1}^4\lambda_1^i \eta_i \right ) {\cal P}^{SG}(p_i) \ .
\ee
The difference with \N=4 supersymmetric Yang-Mills theory is that there is a dimensionful coupling constant. Therefore with extra $\kappa^8$ one can remove the non-polynomial dependence on the transverse momenta: this however, takes place for the 4-point amplitude at the level $\kappa^{2(L-1)}$ which is 4 loops higher than the naive expectation at 3 loops.
For every extra leg in \N=8 one has to add an extra 2 transverse momenta, therefore for the $n$-point amplitude the delay of the UV divergences is increasing at least to the level $L=n+ 3$. This is again a simple analogy with the ``scattering of light-by-light'' concept, since every leg should depend on curvature  and not on the gravitational field $h_{11} \pm i h_{12}$. 

We would like to stress here that the breaking of dynamical supersymmetry means also the breaking of the gauge symmetry, the success of  dynamical supersymmetry  implies the success of the gauge symmetry. However, it may not work the other way around: if we would only 
secure the gauge symmetry, it may be insufficient to claim a success of the dynamical supersymmetry in the supergraphs. Under dynamical supersymmetry all fields in the multiplet transform under the linearized symmetry. Meanwhile, spinors and scalars do not transform under the linearized gauge transformations, for them there is no reason to identify the dynamical supersymmetry with the compensating gauge transformation. This may explain why preserving only the linearized gauge symmetry in the covariant  analysis of the counterterms may be insufficient to see why, for example, the 3, 4, 5, 6-loop divergences of \N=8 supergravity  should be absent according to the light-cone supergraphs analysis. Thus, one should view the analogy with the 
``scattering of light-by-light'' concept in a more general context than just a manifestation of the familiar gauge symmetry.


\section{Discussion}

In conclusion, we have explained here the UV prediction for maximal supersymmetric QFT's, \N=4 supersymmetric Yang-Mills theory and \N=8 supergravity,  by  an  analogy to the ``scattering of light-by-light' effect in the light-cone supergraph method.  The prediction for the absence of \N=8 UV divergences until 7 loops  follows in a simple way from the properly generalized to supergraphs ``scattering of light-by-light'' effect. 

It is also interesting to ask about the possible influence of the \E \, symmetry recently studied in \cite{Brink:2008qc} on UV divergences in higher loops. The rule $L=n+3$ for the delay of the divergences in the $n$-point amplitudes may be combined with \E \, symmetry. The chiral light-cone superfield $\phi(x, \theta)$ transforms non-linearly under \E \, symmetry \cite{Brink:2008qc} and the transformation involves an infinite power 
of fields.  Therefore, it has been proposed in \cite{Kallosh:2009db} that the non-linear nature of this symmetry may require that at any given loop order all $n$-point amplitudes have to be divergent to support a valid counterterm.  This would contradict to the rule $L=n+3$ since for any $L$ there will be some $n$ for which the amplitude is UV finite. So, the hope was expressed in \cite{Kallosh:2009db} that perhaps  \E \, symmetry may  lead to the all loop perturbative  UV  finiteness of \N=8 supergravity.  

More studies will be necessary to clearly understand how the requirement of the non-linear symmetries will affect the UV predictions of the light-cone supergraph method in which so far we used only the linearized symmetries  of the asymptotic fields.

\section*{Acknowledgments}
We are grateful to  L. Brink, D. Freedman, H. Elvang, S. Shenker and L. Susskind  for discussions. 
The work of RK is
supported by the NSF grant 0756174. PR is partially supported by the Department of Energy Grant No. DE-FG02-97ER41029.


\end{document}